\documentclass{iopart}
\usepackage{graphicx}  
\begin{document}

\title[The NuMI Facility]{The NuMI Neutrino Beam and Potential 
for an Off-Axis Experiment}

\author{Sacha E. Kopp\footnote[3]{e-mail address:  kopp@hep.utexas.edu}}
\address{Department of Physics, University of Texas, Austin, TX 78712 U.S.A.}

\begin{abstract}
The Neutrinos at the Main Injector (NuMI) facility at Fermilab is under
construction and due to begin operations in late 2004.  NuMI will deliver
an intense $\nu_{\mu}$ beam of variable energy 2-20~GeV directed into
the Earth at 58~mrad.  Several aspects of the design are reviewed, and 
potential limitations to the ultimate neutrino flux are described.  In 
addition, potential measurements of neutrino mixing properties are described.
\end{abstract}


%\submitto{\JPG}

%\maketitle

\section{Introduction}
\label{intro}

Within the context of 3-flavor neutrino flavor mixing \cite{mns}, the 
atmospheric neutrino experiments \cite{atmosnu} 
have observed $\nu_{\mu} \rightarrow \nu_{\tau}$
oscillations with $|\Delta m_{23}^2|=(2-4)\times10^{-3}$~eV$^2$ and 
$\theta_{23} \approx \pi/4$, while solar neutrino experiments\cite{solarexp}
indicate $\Delta m_{21}^2=(2-10)\times10^{-5}$~eV$^2$ and 
$\theta_{12}\approx\pi/6$.  Unlike the quark mixing matrix, neutrino mixing 
appears near-maximal.  Only upper bounds \cite{nue} exist on the process 
$\nu_{\mu} \leftrightarrow \nu_e$ at the atmospheric $\Delta m^2$.
\footnote[1]{The LSND experiment has published evidence\cite{lsnd} for 
$\overline{\nu}_{\mu} \rightarrow \overline{\nu}_e$ at  
$\Delta m^2\sim0.1$~eV$^2$ which, if confirmed, signficantly modifies
the 3-flavor mixing scheme above.}  If the angle $\theta_{13}$ is not
zero, and in fact is reasonably large, then the potential exists
at conventional long-baseline neutrino beams to measure
this rare process and to use it to observe CP violation in the lepton
sector and determine the mass hierarchy of neutrinos \cite{cpviol}.

\begin{figure}
\includegraphics*[width=60mm]{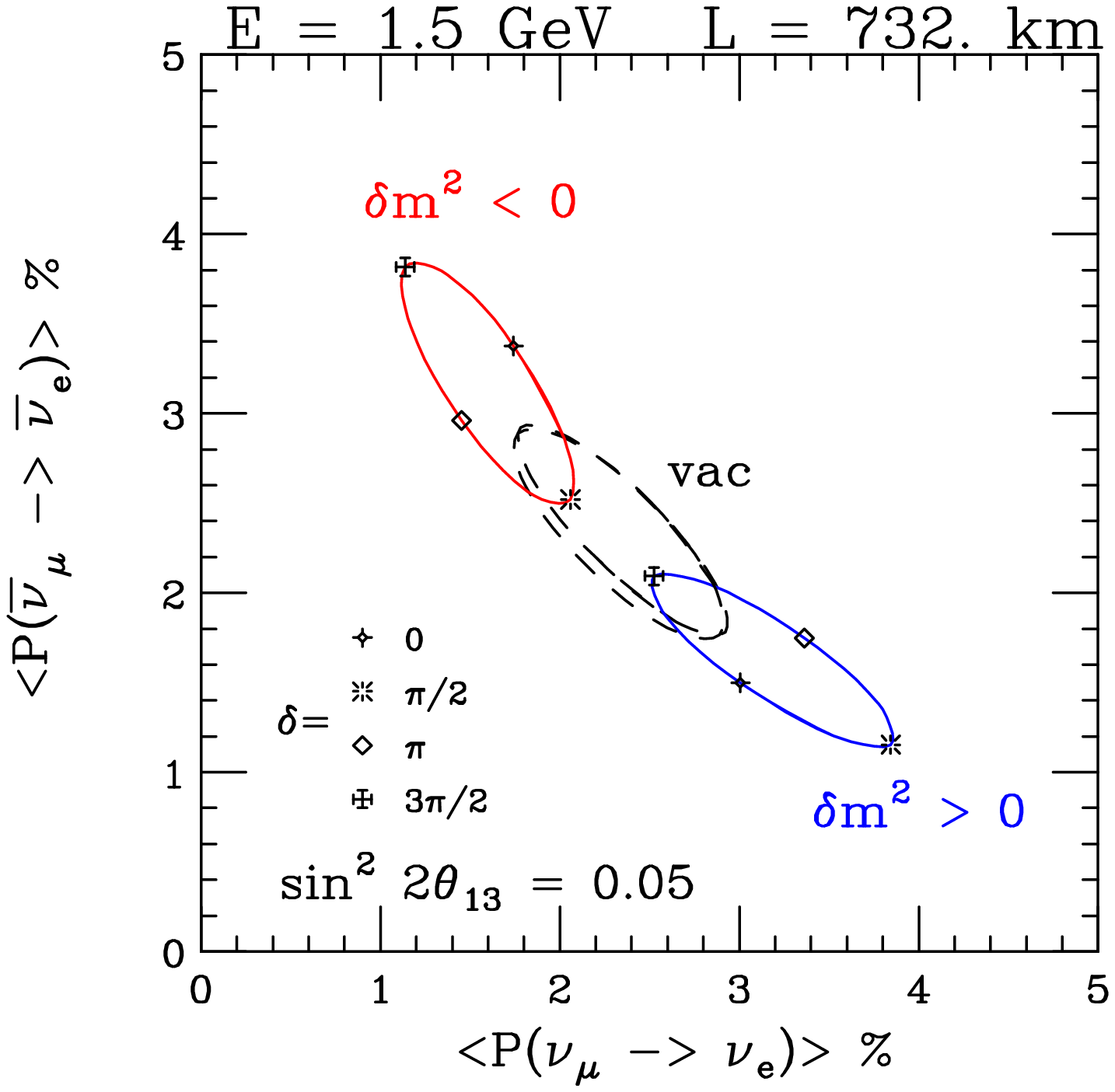}
\hfill
\includegraphics*[width=62mm]{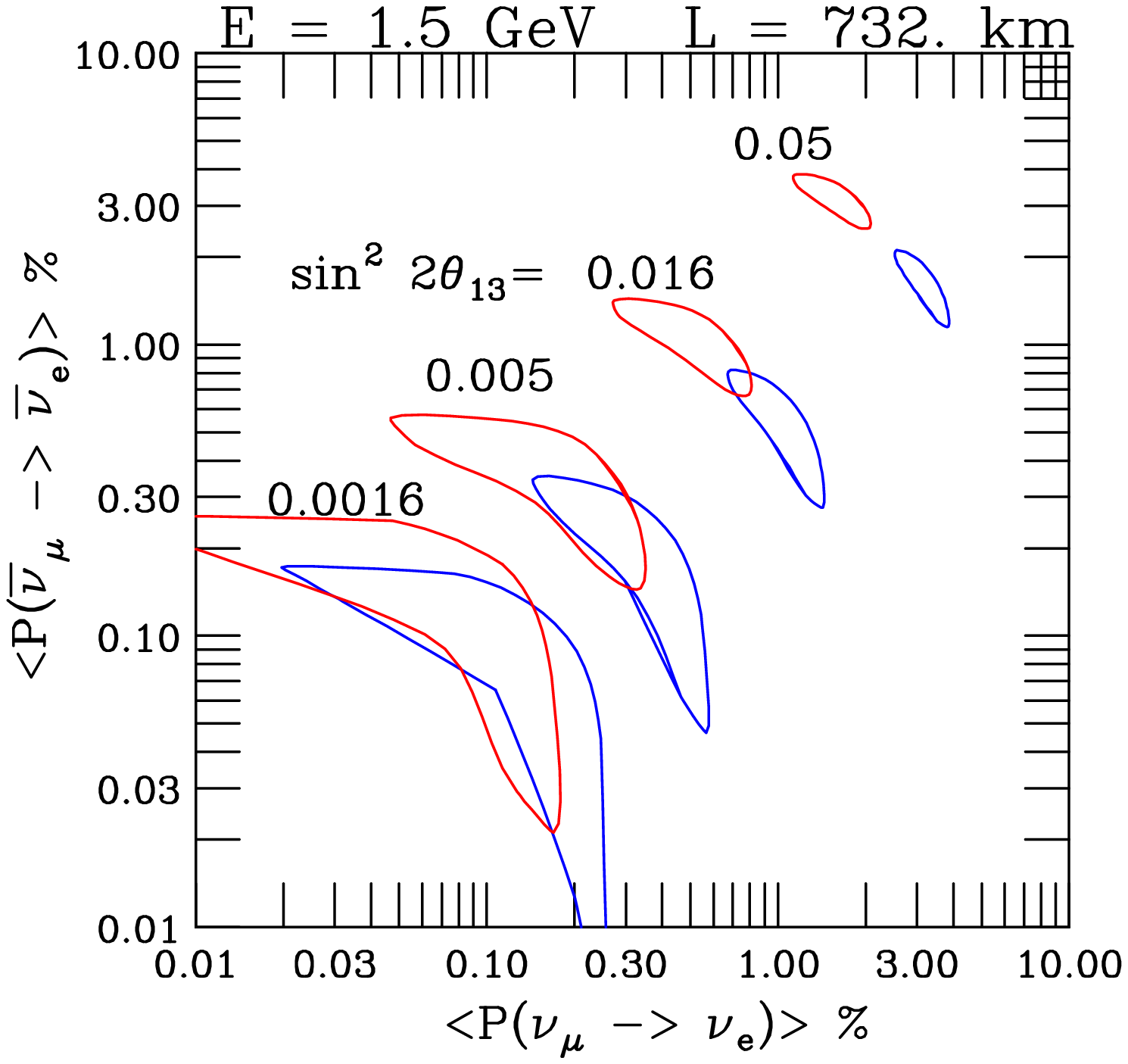}
\caption{\label{ellipse} The expected oscillation probability for 
$\nu_{\mu}\rightarrow \nu_e$ and 
$\overline{\nu}_{\mu} \rightarrow \overline{\nu}_e$ evaluated at 
$L/E$=500~km/GeV and 
$|\Delta m_{31}^2|=3\times10^{-3}$~eV$^2$, $\sin^2 2\theta_{23}=1.0$, 
$\Delta m_{21}^2=+5\times10^{-5}$~eV$^2$, $\sin^2 2\theta_{12}=0.8$, 
and a constant matter density $\rho=3.0$~g/cm$^{-3}$.  
(Left)
evalated ignoring matter effects (black ellipse) and including matter effects
for positive (blue) and negative (red) $\Delta m_{31}^2$.  The values
of $\sin^2 2\theta_{13}$ and $\delta$ are indicated.  (Right) The same
bi-probability plots for several possible values of $\sin^2 2\theta_{13}$.}
\end{figure}

To first order the appearance of $\nu_e$ in a pure $\nu_{\mu}$ beam in vacuum is given by 
$P(\nu_{\mu}\rightarrow \nu_e)= \sin^2\theta_{23}\sin^2 2\theta_{13}
\sin^2(1.27 L \Delta m^2_{23}/E_{\nu})$, which is approximately 
$\frac{1}{2}\sin^2 2\theta_{13}$ at the atmospheric $L/E_{\nu}$.  Given the
Chooz bounds \cite{nue} an experiment to observe this 
transition must be sensitive at the 1\% level.  This transition is modified
in matter by a factor $(1 \pm 2E_{\nu}/E_R)$, where $E_R\sim 13$~GeV is the
resonance energy in the earth's crust at the atmospheric $\Delta m^2$.  
Neutrino (antineutrino) transitions are enhanced (depreciated) by this factor 
under the normal mass hierarchy, while the opposite
occurs for an inverted mass hierarchy.
This transition is also modified by a factor which depends on the phase
$\delta$ of the neutrino mixing matrix, potentially yielding the 
opportunity to
observe a CP-violating difference between 
$\nu_{\mu}\rightarrow \nu_e$ and 
$\overline{\nu}_{\mu} \rightarrow \overline{\nu}_e$ transitions.  
Figure~\ref{ellipse} shows the probabilities for these transitions at
an $L/E_{\nu}$ accessible with the NuMI beam.

\begin{figure}[t]
\includegraphics*[width=60mm]{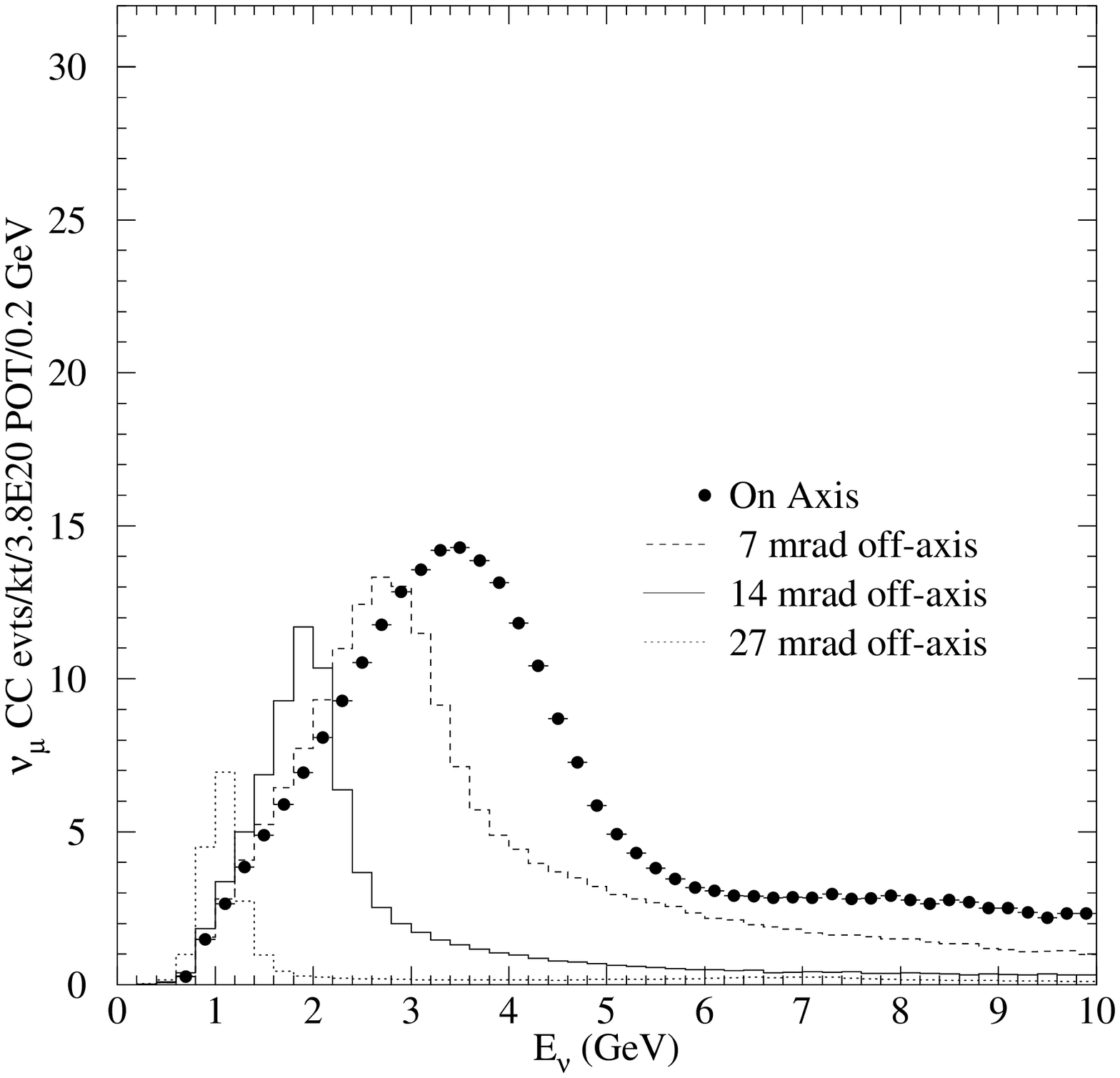}
\hfill
\includegraphics*[width=60mm]{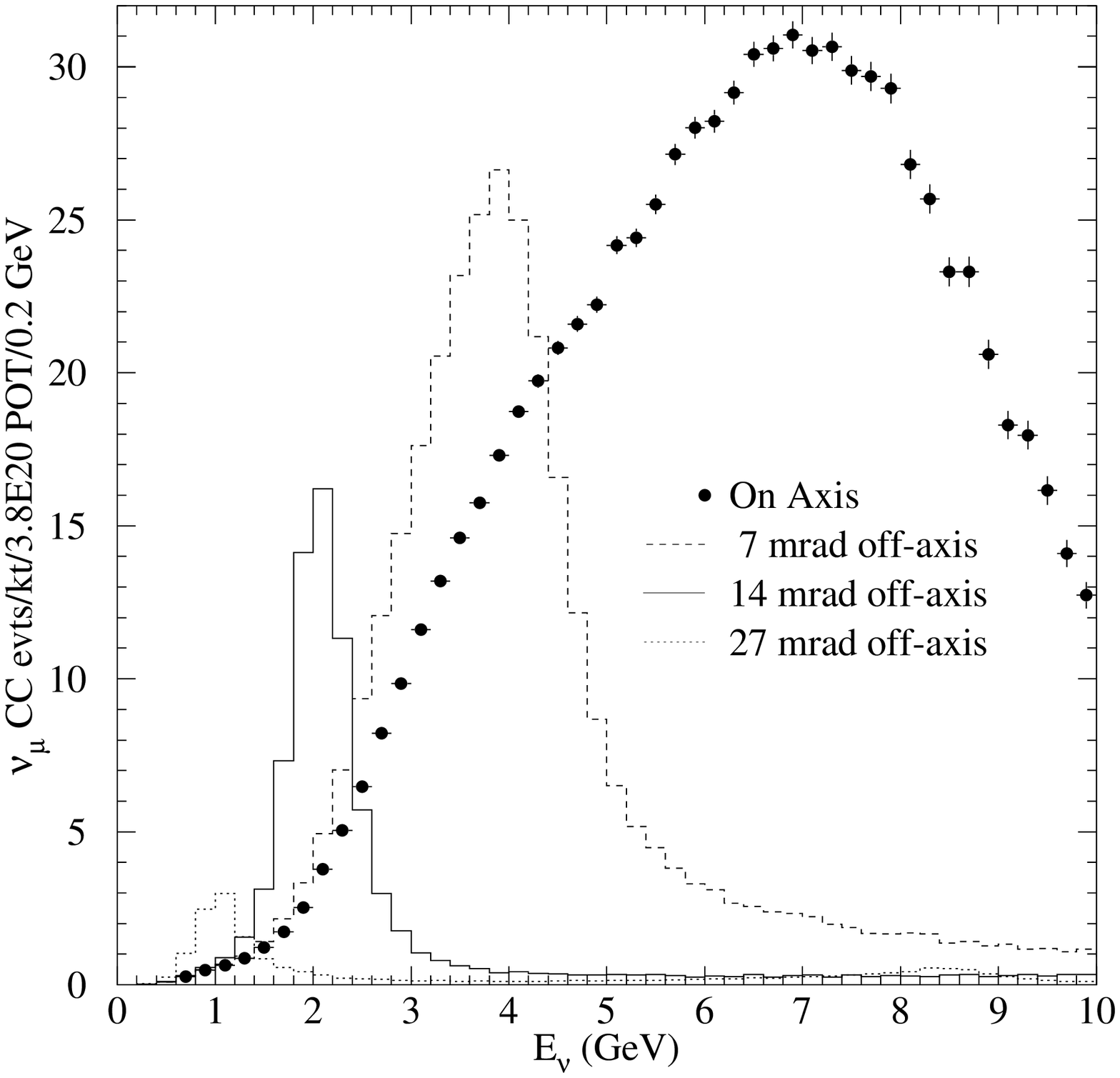}
\caption{\label{spectra}Energy spectra of charged current events expected 
at a far detector location 735~km from Fermilab at various off-axis angles 
for the NuMI low-energy beam setting (left) and the medium-energy setting
(right).}
\end{figure}

\begin{figure}[b]
 \includegraphics*[width=60mm]{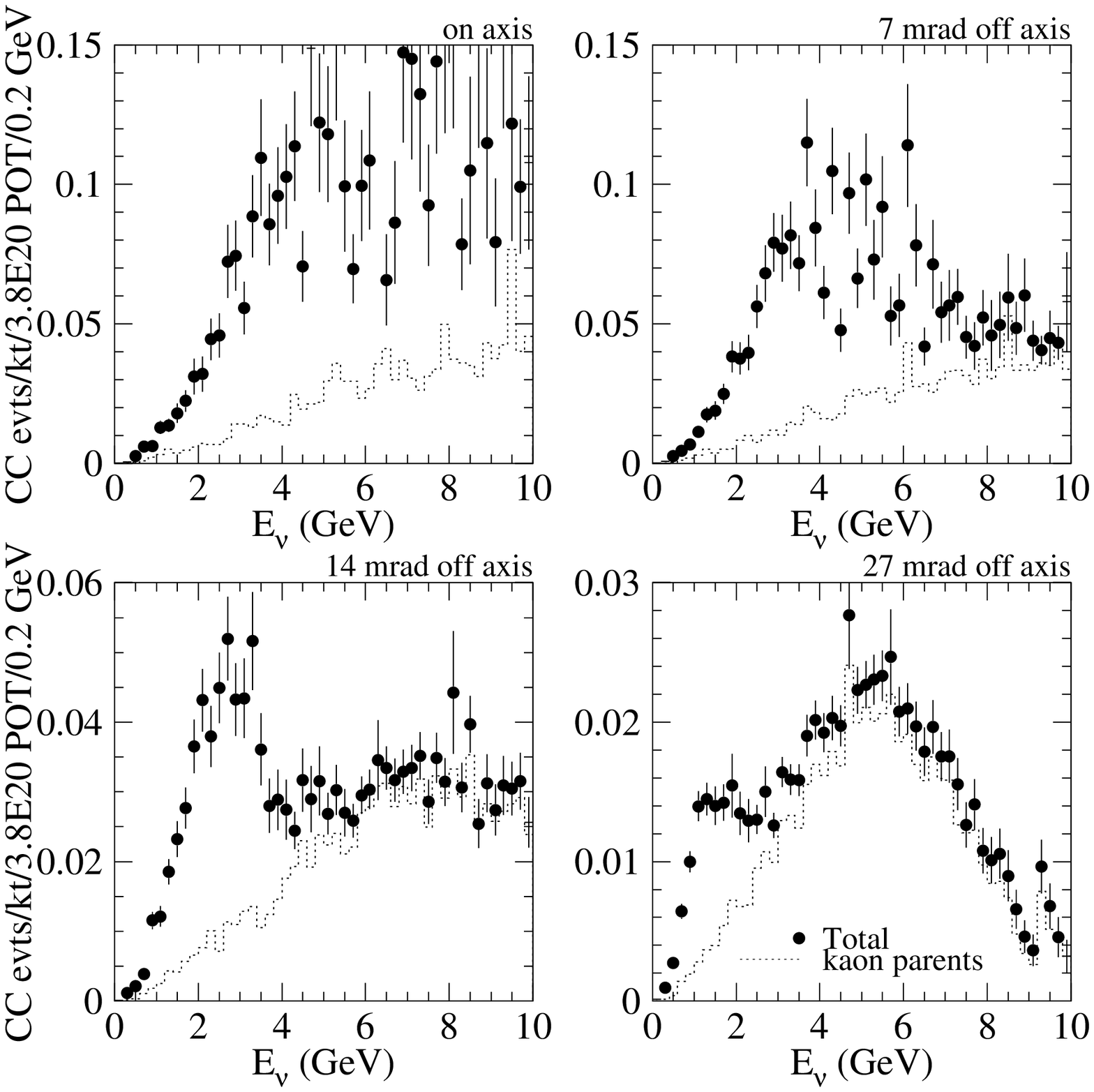}
 \includegraphics*[width=68mm]{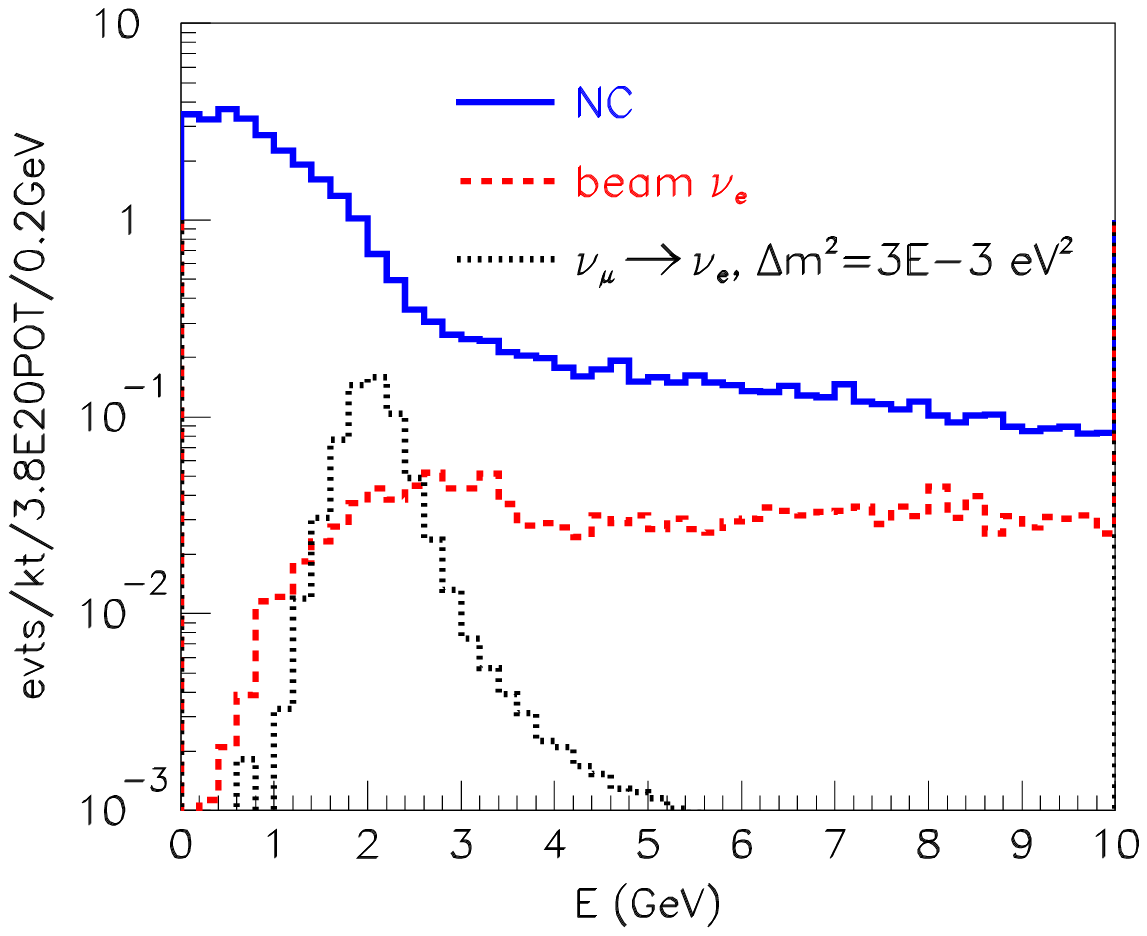}
 \caption{\label{bckgd}
 (Left)  Spectrum of $\nu_e$ events expected at a far detector
 735~km from Fermilab for several off-axis angles.  The $\nu_e$ rates are 
 plotted in total (points) and the component from $K\rightarrow \pi e\nu_e$ 
 decays.  (Right) Spectrum of potential $\nu_{\mu} \rightarrow \nu_e$ 
 oscillation signal with $|U_{e3}|^2=0.01$ (black histogram) and for two 
 potential $\nu_e$ backgrounds:  real $\nu_e$ from muon and $K_{e3}$ decays 
 (red histogram) and the total visible (hadronic) energy from $\nu_{\mu}$ 
 neutral current events before any rejection criteria are applied (blue
 histogram).}
 \end{figure}

Figure~\ref{spectra} shows the neutrino spectra possible in NuMI\cite{numi}, 
both for the initial MINOS experiment\cite{minos} on-axis, 
and at three possible off-axis 
angles, where the resulting neutrino energy from off-axis pion decays
becomes rather narrowly peaked and near-independent of pion energy 
due to two-body decay kinematics\cite{bnloff}.  At 14~mrad, the peak is near
oscillation maximum, and the intrinsic $\nu_e$ from the beam is of order
0.4\% under the peak (see Figure~\ref{bckgd}), coming primarily from muon decays.  
Of greater magnitude is the potential background to a $\nu_e$ appearance search 
from neutral current $\nu_{\mu}$ interactions, which leave a small hadronic
visible energy in the detector (see Figure~\ref{bckgd}).  A study of detector 
designs for an off-axis experiment will be required to assess how frequently 
these are misidentified as charged current $\nu_e$ interactions.  

The event rate spectra in Figures~\ref{spectra} and \ref{bckgd} are possible
with the baseline NuMI beam design, so that if $\sin^2 2\theta_{13}=0.1$ 
(at the Chooz limit), a 20~kt NuMI experiment at 712~km with similar $\nu_e$
efficiency and NC 
rejection as SuperK might observe 86 oscillated $\nu_e$'s in 5 years, with a 
background of 10 NC's and 10 intrinsic beam $\nu_e$.  For JHF \cite{jhf}, 
the similar numbers are 123 signal, 11 beam $\nu_e$, and 11 NC's in their
Phase I.  Note that the lower proton beam power at NuMI is mitigated by the
higher cross sections for 1.8~GeV neutrinos.  
To measure matter effects (hence the sign of $\Delta m_{13}^2$)
or the phase $\delta$ will require running in $\overline{\nu}_{\mu}$
mode and additional upgrades to the proton beam intensity from the Main
Injector \cite{protondriver} because of the lower antineutrino event rates.
Both these measurements benefit from longer baseline
distances, again motivating the need for proton intensity upgrades and 
furthermore larger detectors.

%\begin{figure}
%\includegraphics*[width=60mm]{oafluxepi.eps}
%\hfill
%\includegraphics*[width=60mm]{oaepienu.eps}
%\caption{\label{offaxis}Left:  The neutrino flux from a parent pion of energy
%$E_{\pi}$ for several decay angles $\theta$ between the pion and neutrino 
%directions.  The flux is evalutated at a distance of 735~km.  Right:  The
%energy of the daughter neutrino {\it vs} the parent pion energy for several
%decay angles.}
%\end{figure}

\section{The NuMI Beamline}
\label{numi}

NuMI is a tertiary beam resulting from the decays
of pion and kaon secondaries produced in the NuMI target. Protons of 
120~GeV are fast-extracted (spill duration 8.6~$\mu$sec)
from the Main Injector (MI) accelerator and bent downward by 58~mrad toward Soudan, 
MN.  The beamline is designed to accept $3.8\times 10^{13}$~protons per pulse 
(ppp).  The spill repetition rate is 0.45~Hz, giving $4 \times 10^{20}$~protons
on target per year.

The Main Injector is fed up to 6 batches from the Booster accelerator, of
which 5 batches are extracted to NuMI.  
It is likely that when Main Injector operates at full intensity in 
multi-batch mode it will have $\Delta p/p~=~0.003$, and emittances 
up to 40$\pi$~mm-mrad.  Initiatives to produce higher intensity beam 
from the Main Injector without constructing a proton driver upgrade, 
including barrier RF stacking of more than 6 
Booster batches, could increase the beam intensity by a factor of 1.5
at the expense of some emittance growth.

The NuMI optics will maintain losses below $10^{-5}$.  
A recent redesign of the primary beamline  
replaced a long free drift region with a FODO lattice,
increasing the momentum acceptance to 
$\Delta p/p~=~0.0038$ at 40$\pi$~mm-mr.  
Injection errors of $\sim$1~mm 
lead to targeting errors of $\sim$0.5~mm.
The beamline has 21 BPM's of 50~$\mu$m accuracy, 
19 dipole correctors, and 10 retractable 
secondary emission foil profile+halo monitors.

The primary beam is focused onto a graphite production target \cite{target}
of 6.4$\times$ 20$\times$~940~mm$^3$ segmented longitudinally into
47 fins.  The target is water cooled via stainless steel lines at the top
and bottom of each fin and is contained in an
aluminum vacuum can with beryllium windows.  It is electrically 
isolated so it can be read out as a Budal monitor \cite{budal}.  
The target has a safety
factor of about 1.6 for the fatigue lifetime of 10$^7$ pulses (1 NuMI year) 
given the calculated dynamic stress of 
$4\times~10^{13}$~protons/pulse and 1~mm spot size.  
A prototype target was tested in the Main Injector at peak energy densities
exceeding that expected in NuMI.  Studies indicate that the
existing NuMI target could withstand up to a 1~MW proton beam if the beam
spot size is increased from 1~mm to 3~mm \cite{mikhail}.  

The particles produced in the target are focused by two magnetic 'horns' 
\cite{horn}.  The 200~kA peak current produces
a toroidal field which sign- and 
momentum-selects the particles from the target.  The relative placement of the 
two horns and the target optimizes the momentum focus for pions
to a particular momentum,
hence the peak neutrino beam energy.  The neutrino spectra from two energy 
settings, called the 'low energy' (LE) and 'medium energy' (ME) beams, are 
shown in Figure~\ref{spectra}.  To fine-tune the beam energy,
the target is mounted on a rail-drive system with 2.5~m of travel along the
beam direction, permitting remote change of the beam energy without 
accessing the horns and target \cite{variablebeam}.  Measurements on
a horn prototype show the expected $1/r$ fall-off of the field to within a 
percent.  The horns are designed to withstand 10$^7$
pulses (1 NuMI year), and tests of the prototype horn have so far 
achieved this.  

The particles are focused forward by the horns into a 675~m long, 2~m
diameter steel pipe evacuated to $\sim$~0.1~Torr.  This length is approximately
the decay length of a 10~GeV pion.  The entrance window to the decay volume
is a spherical bell-shaped steel window 1.8~cm in thickness, with a 1.5~mm
thick aluminum window 1~m in diameter at its center where 95\% of the 
entering pions traverse.  The decay volume is surrounded by 
2.5-3.5~m of concrete shielding.  Twelve water cooling lines around the
exterior of the decay pipe remove the 150~kW of beam heating.    
Earlier plans to instrument the
decay volume with a current-carrying wire (called the 'Hadron Hose' 
\cite{hose}) which would provide a toroidal field that continuously 
focuses pions along the decay pipe length, have been
abandoned due to budget constraints.

At the end of the decay volume is a beam absorber consisting of a 
1.2$\times$1.2$\times$2.4~m$^3$ water-cooled aluminum core, a 1~m thick
layer of steel blocks surrounding the core, followed by a 1.5~m thick
layer of concrete blocks.  The core absorbs 65~kW of beam power, but is
capable of taking the full proton beam power of 400~kW for up to an hour
in the event of a mistargeting. In the event of a proton intensity upgrade
the core would require no modification, but the steel blocks 
might require cooling.  

Ionization chambers are used to monitor the secondary beam.  An array
is located immediately upstream of the absorber, as well as at three 
muon 'pits', one downstream of the absorber, one after 8~m of rock, and a
third after an additional 12~m of rock.  These chambers 
monitor the remnant hadrons at the end of the decay pipe, as well as
the tertiary muons from $\pi$ and $K$ decays.  When the beam is tuned to the 
medium energy configuration, the pointing accuracy of the muon stations
can align the neutrino beam direction to approximately 50~$\mu$Radians
in one spill.  Beam tests\cite{atf,booster} of these chambers
indicate an order of magnitude safety factor in particle flux over the 
10$^9$/cm$^2$/spill expected in NuMI before space charge buildup affects 
their operation.

\section{Conclusions}

The NuMI beam will turn on in 2004, with the first experiment, MINOS, ready 
to take data.  Beyond MINOS, a potential physics program is to exploit the 
intense NuMI beam in an off-axis experiment, where sensitivities  to 
$\nu_{\mu} \rightarrow \nu_e$ at the percent level would allow competitive 
measurements to see if this rare transition is large enough for  CP violation 
studies.  Additionally, the NuMI physics program is unique and complements that 
at the JHF in that matter effects and the neutrino mass hierarchy may be 
explicitely studied along with the potential CP-violating phase $\delta$.

%
%This creates an acknowledgement section in the IOP style
%
\ack

It is a pleasure to thank my NuMI/MINOS colleagues, particularly 
Jim Hylen, Bruce Baller, Dixon Bogert, Bob Ducar, Dave Pushka, Adam Para, Debbie Harris, 
Mark Messier, and Karol Lang.  Many of the oscillations studies
grew out of a Fermilab study group and a Letter of Intent written to Fermilab this
year.  I thank
Debbie Harris for the invitation to speak at this workshop.  Support of the
U.S. Department of Energy DE-FG03-93ER40757 
and the Fondren Family Foundation are acknowledged.

\end{document}